\documentclass[%
 aip,jcp,
 amsmath,amssymb,
 reprint,
]{revtex4-2}

\usepackage{graphicx}
\usepackage{dcolumn}
\usepackage{bm}

\usepackage[utf8]{inputenc}
\usepackage[T1]{fontenc}
\usepackage{mathptmx}
\usepackage{listings}
\usepackage{rotating} 

\usepackage{color}
\usepackage{dcolumn} 
\usepackage{bm} 
\usepackage{graphicx}
\usepackage{multirow} 
\usepackage{pifont} 
\usepackage{epsfig}
\usepackage{amsmath} 
\usepackage{subfigure}
\usepackage{float}
\usepackage{booktabs}
\usepackage{tabularx}
\usepackage{natbib}
\usepackage{gensymb}
\usepackage{rotating}
\usepackage{threeparttable}
\usepackage{comment}
\usepackage[normalem]{ulem}
\usepackage[hidelinks]{hyperref} 

\begin{document}

\title{Assessing the Effects of Orbital Relaxation and the Coherent-State Transformation in Quantum Electrodynamics Density Functional and Coupled-Cluster Theories}

\author{Marcus D. Liebenthal}
\affiliation{
             Department of Chemistry and Biochemistry,
             Florida State University,
             Tallahassee, FL 32306-4390, USA}
  
\author{Nam Vu}
\affiliation{
             Department of Chemistry and Biochemistry,
             Florida State University,
             Tallahassee, FL 32306-4390, USA}          
             
\author{A. Eugene DePrince III}
\email{adeprince@fsu.edu}
\affiliation{
             Department of Chemistry and Biochemistry,
             Florida State University,
             Tallahassee, FL 32306-4390, USA}



\begin{abstract}

Cavity quantum electrodynamics (QED) generalizations of time-dependent (TD) density functional theory (DFT) and equation-of-motion (EOM) coupled-cluster (CC) theory are used to model small molecules strongly coupled to optical cavity modes.  We consider two types of calculations. In the first approach (termed ``relaxed''), {\color{black} we use a coherent-state-transformed Hamiltonian within the ground- and excited-state portions of the calculations, and cavity-induced orbital relaxation effects are included at the mean-field level.} This procedure guarantees {\color{black}that the energy is origin invariant} in post-self-consistent-field calculations.
In the second approach (termed ``unrelaxed''), we ignore the coherent-state transformation and the associated orbital relaxation effects. In this case, ground-state unrelaxed QED-CC calculations pick up a modest origin dependence but otherwise reproduce relaxed QED-CC results within the coherent-state basis. On the other hand, a severe origin dependence manifests in ground-state unrelaxed QED mean-field energies. For excitation energies computed at experimentally realizable coupling strengths, relaxed and unrelaxed QED-EOM-CC results are similar, while significant differences emerge for unrelaxed and relaxed QED-TDDFT. First, QED-EOM-CC and relaxed QED-TDDFT both predict that electronic states that are not resonant with the cavity mode are nonetheless perturbed by the cavity. Unrelaxed QED-TDDFT, on the other hand, fails to capture this effect. Second, in the limit of large coupling strengths, relaxed QED-TDDFT tends to overestimate Rabi splittings, while unrelaxed QED-TDDFT underestimates them, given splittings from relaxed QED-EOM-CC as a reference, and relaxed QED-TDDFT generally does the better job of reproducing the QED-EOM-CC results.

\end{abstract}

\maketitle

\section{Introduction}

Chemical applications of strong light-matter interactions facilitated by optical cavities have garnered a great deal of attention in recent years.\cite{Ebbesen16_2403,Narang18_1479,YuenZhou18_6325,Nori19_19} This interest has been driven by experimental studies offering evidence that strong light-matter coupling and polariton formation can be leveraged in chemical contexts, \cite{Bucksbaum15_164003, Shegai18_eaas9552, KenaCohen19_eaax4482, Nori19_19} such as for catalyzing/inhibiting reactions\cite{Ebbesen12_1592, Ebbesen16_11462, Borjesson18_2273, George19_10635, Shalabney21} or controlling reaction selectivity.\cite{Ebbesen19_615} Moreover, a large number of computational studies have predicted a range of phenomena that are relevant to chemistry.\cite{Muschik11_839, Narang18_1479, Nori19_19, Corni18_4688, Zhou19_4685, Feist19_131, Flick22_4995, DePrince22_9303} Predictive electronic/polaritonic structure methods will be crucial for discovering general design principles for cavity-mediated chemistry; as a result, substantial effort has been dedicated to the generalization of familiar tools in quantum chemistry for the polaritonic problem. 

Proposed cavity quantum electrodynamics (QED) models incorporating an {\em ab initio} treatment of molecular degrees of freedom have largely taken one of two complementary approaches. First, given the success that density functional theory (DFT) has seen in standard quantum chemical applications, a large body of work has considered quantum electrodynamical generalizations of DFT\cite{Rubio14_012508, Rubio17_3026, Rubio17_113036, Rubio21_e2110464118} and time-dependent DFT (TDDFT).\cite{Tokatly13_233001, Tokatly18_235123, Rubio20_508, Narang20_094116, Narang21_104109, Shao21_064107,Shao22_124104, Varga22_194106, DePrince22_9303} QED-DFT and QED-TDDFT provide access to orbital-specific quantities that cannot be directly probed with model Hamiltonians;\cite{Cummings63_89,Cummings68_379} because they inherit the favorable computational scaling of conventional DFT and TDDFT, these methods can be applied to large cavity-embedded molecules or collections of molecules. At the same time, the well-known issues that plague DFT\cite{Yang08_792} and the small number of exchange--correlation functionals that have been developed for the polaritonic problem\cite{Rubio15_093001,Rubio18_992,Flick22_143201} have inspired others to pursue correlated wave-function-based approaches to polaritonic structure,\cite{Koch20_041043,Manby20_023262,DePrince21_094112,DePrince22_054105,Flick22_4995,Koch21_094113,Flick21_9100,Koch22_2203,Rubio22_094101,Koch22_chiral,Rubio23_2766,Knowles22_204119, Varga21_273601,Foley22_154103} within formalisms that resemble familiar coupled-cluster (CC)\cite{Cizek66_4256, Paldus71_359, Bartlett09_book, Musial07_291, Bartlett07_291} or configuration interaction (CI) approaches. Like QED-DFT and QED-TDDFT, QED generalizations of correlated wave-function methods can provide insight into subtle cavity-induced changes to electronic structure, while also offering the advantage of systematic improvability.

Straightforward polaritonic generalizations of ground-state CC and equation-of-motion (EOM) CC\cite{Bartlett93_7029,Bartlett12_126,Musial07_291,Krylov08_433} have been put forth in  Ref.~\citenum{Koch20_041043}. The QED-(EOM)-CC formalism developed therein has subsequently been applied in a number of studies (illustrating, for example, how cavity interactions can influence electron ionization/attachment\cite{DePrince21_094112, DePrince22_054105,Flick21_9100, Koch22_2203} reaction rates,\cite{Rubio23_2766, Flick22_4995} and non-bonded interactions\cite{Koch21_094113}), and the family of QED-CC-inspired approaches also continues to grow. QED-(EOM)-CC has been  generalized to make use of non-particle-conserving operators,\cite{DePrince22_054105} to employ unitary cluster operators,\cite{Flick21_9100}  for the description of chiral cavity modes,\cite{Koch22_chiral} and for wave-function-in-DFT embedding protocols.\cite{Rubio22_094101}

As mentioned above, one of the attractive features of QED-based many-body theories such as QED-CC (and QED-EOM-CC) relative to QED-DFT (and QED-TDDFT) is the systematic improvability of the former approach. An equally important but underappreciated aspect of QED-CC is that it is robust against changes to the reference orbitals. This property is inherited from the conventional (non-QED) formulation of CC theory and stems from the presence of the exponentiated single excitation operator, $e^{\hat{T}_1}$, which closely resembles an orbital rotation operator (except that it is not unitary). Indeed, it is well known that energies calculated at the CC with single and double excitations (CCSD)\cite{Bartlett82_1910} level of theory often closely reproduce energies computed using the Bruekner coupled-cluster doubles (BCCD) approach,\cite{Scuseria87_354, Handy89_185, Werner92_112} which variationally optimizes the orbitals for the coupled-cluster doubles wave function. In the context of QED-CC,  $e^{\hat{T}_1}$ should be able to account for orbital relaxation effects induced by the cavity in the underlying QED-HF wave function should one choose to seed a QED-CC calculation with a non-QED Hartree-Fock reference configuration (see Fig.~\ref{FIG:RELAXED_VS_UNRELAXED}); QED-EOM-CC results obtained in either case should then be similar. A related matter derives from the fact that QED-CC calculations are typically carried out using a coherent-state transformed Hamiltonian,\cite{Koch20_041043} which guarantees invariance of the QED-CC energy with respect to the placement of the origin. The QED-CC cluster operator contains an exponentiated boson creation operator term that should be able to mimic the effects of this transformation, and thus we expect the origin dependence of QED-CC to be modest when using a Hamiltonian that has not been transformed to the coherent-state basis. On the other hand, as we demonstrate below, QED-TDDFT results are quite sensitive to whether or not the Kohn-Sham orbitals are allowed to relax in the presence of the cavity and whether the Hamiltonian that enters the QED-TDDFT equations is represented within the coherent-state basis. This point is subtle, yet important, given that a variety of QED-TDDFT prescriptions have been put forth and not all of them account for the presence of the cavity self-consistently.\cite{Shao21_064107, Shao22_124104}

{\color{black}In this work, we examine how cavity-induced changes to the orbitals and the coherent-state transformation affect the energies of cavity-embedded molecules treated at the QED-EOM-CC and QED-TDDFT levels of theory. Before doing so, we present the theory underlying relaxed and unrelaxed versions of these approaches, which differ in the treatment of the cavity at the mean-field level. The details of our calculations are then provided in the Computational Details, and numerical studies exploring the robustness of QED-TDDFT and QED-EOM-CC to the description of cavity effects at the mean-field level can be found in the Results and Discussion. Lastly, we conclude with a summary of the outcomes from our numerical studies.}

\section{Theory}

\label{SEC:THEORY}

In this Section, we outline some key details of the QED-EOM-CC and QED-TDDFT approaches. Both of these methods model the physics of a cavity-embedded molecular system using the Pauli--Fierz Hamiltonian,\cite{Spohn04_book,Rubio18_0118} which we represent in the length gauge and within the dipole and Born-Oppenheimer approximations. For a single-mode cavity, this
Hamiltonian takes the form
\begin{eqnarray}
    \label{EQN:PFH}
    \hat{H}_{\rm PF} &=& \hat{H}_{\rm e} + \omega_{\rm cav} \hat{b}^\dagger \hat{b} - \sqrt{\frac{\omega_{\rm cav}}{2}} \left({\bm \lambda} \cdot {\bm \hat{\mu}}\right)\left(\hat{b}^\dagger +\hat{b}\right) \nonumber + \frac{1}{2} ({\bm \lambda} \cdot {\bm \hat{\mu}} )^2\\
\end{eqnarray}
Here, $\hat{H}_{\rm e}$ and $\omega_{\rm cav}\hat{b}^\dagger\hat{b}$ are the Hamiltonians for the isolated many-electron system and the cavity mode, respectively; $\omega_{\rm cav}$ is the frequency of the cavity mode, and $\hat{b}^\dagger$ ($\hat{b}$) is a bosonic creation (annihilation) operator. The third term in Eq.~\ref{EQN:PFH} describes the coupling between the molecular degrees of freedom and the cavity mode, which is parametrized by the coupling strength, $\bm \lambda$; the symbol $\hat{\bm \mu}$ represents the molecular dipole operator. The fourth term is the dipole self-energy contribution. In the single-molecule coupling limit, the coupling strength is related to the effective mode volume, $V_{\rm eff}$, as 
\begin{equation}
    \label{EQN:lambda}
    {\bm \lambda} = \lambda \hat{\bm e} = \sqrt{\frac{4\pi}{V_{\rm eff}}} \hat{\bm e}
\end{equation}
where $\hat{\bm e}$ is a unit vector pointing along the cavity mode polarization axis. 

\subsection{Cavity QED Coupled-Cluster Theory for Ground and Excited States}

\subsubsection{Cavity QED Hartree-Fock Theory}

The cavity QED Hartree-Fock (HF) wave function
is a product of a Slater determinant of molecular spin orbitals, $\left|0^{\rm e}\right\rangle$, and a zero-photon state, $\left|0^{\rm p}\right\rangle$. 
Following Ref.~\citenum{Koch20_041043}, the zero-photon state can be exactly represented using the
coherent-state (CS) transformation
\begin{eqnarray}
    \left|0^{\rm p}\right\rangle &=& \hat{U}_{\rm CS} | 0 \rangle = \exp \left(z \hat{b}^{\dagger}-z^{\ast} \hat{b}\right) \left|0\right\rangle
    \label{EQN:COHERENT_TRANSFORM}
\end{eqnarray}
Here, $|0\rangle$ is the photon vacuum,  and
\begin{equation}
\label{EQN:Z}
    z = -\frac{{\bm \lambda} \cdot\langle {\bm \mu}\rangle}{\sqrt{2 \omega}}
\end{equation}
The symbol $\langle {\bm \mu}\rangle$ represents the expectation value of the molecular dipole operator with respect to $|0^{\rm e}\rangle$. 
{\color{black}One} can use $\hat{U}_{\rm CS}$ to transform the Hamiltonian to the coherent-state basis to give
\begin{eqnarray}
    \label{EQN:PFH_COHERENT}
    \hat{H}_{\rm CS} &=& \hat{U}_{\rm CS}^\dagger\hat{H}\hat{U}_{\rm CS} \nonumber \\
    &=& \hat{H}_{\rm e} + \omega_{\rm cav} \hat{b}^\dagger \hat{b} - \sqrt{\frac{\omega_{\rm cav}}{2}} \left({\bm \lambda} \cdot \left[{\bm \mu}-\left\langle {\bm \mu} \right\rangle\right]\right)\left(\hat{b}^\dagger +\hat{b}\right) \nonumber \\
    &+& \frac{1}{2} ({\bm \lambda} \cdot \left[{\bm \mu}-\left\langle {\bm \mu} \right\rangle\right] )^2
\end{eqnarray}
In the coherent-state basis, the QED-HF wave function has the simple form
\begin{equation}
\label{EQN:QED_HF_WFN}
|\Phi_{0} \rangle = |0^{\rm e}\rangle \otimes |0\rangle
\end{equation}
and  $\left|0^{\rm e}\right\rangle$ can be determined via a standard SCF procedure using $\hat{H}_{\rm CS}$ after integrating out the photon degrees of freedom, {\em i.e.}, 
\begin{equation} 
\label{EQN:PFH_COHERENT_SCF}
    \langle 0 | \hat{H}_{\rm CS} | 0 \rangle = \hat{H}_{\rm e} + \frac{1}{2} ({\bm \lambda} \cdot \left[{\bm \mu}-\left\langle {\bm \mu} \right\rangle\right] )^2
\end{equation}
We refer to a QED-HF wave function determined in this way as ``relaxed,'' in the sense that the electronic spin orbitals account for the presence of the cavity (through the dipole self energy term in Eq.~\ref{EQN:PFH_COHERENT_SCF}); the relaxed mean-field energy is the expectation value of Eq.~\ref{EQN:PFH_COHERENT} with respect to Eq.~\ref{EQN:QED_HF_WFN}. On the other hand, an ``unrelaxed'' QED-HF wave function has the same form (Eq.~\ref{EQN:QED_HF_WFN}), but $|0^{\rm e}\rangle$ is instead determined from an SCF procedure that considers only $\hat{H}_{\rm e}$. The unrelaxed mean-field energy is the expectation value of Eq.~\ref{EQN:PFH} with respect to Eq.~\ref{EQN:QED_HF_WFN}.

\subsubsection{Ground-state QED-CC theory}

\begin{figure}[!htpb]
\includegraphics[width=\linewidth]{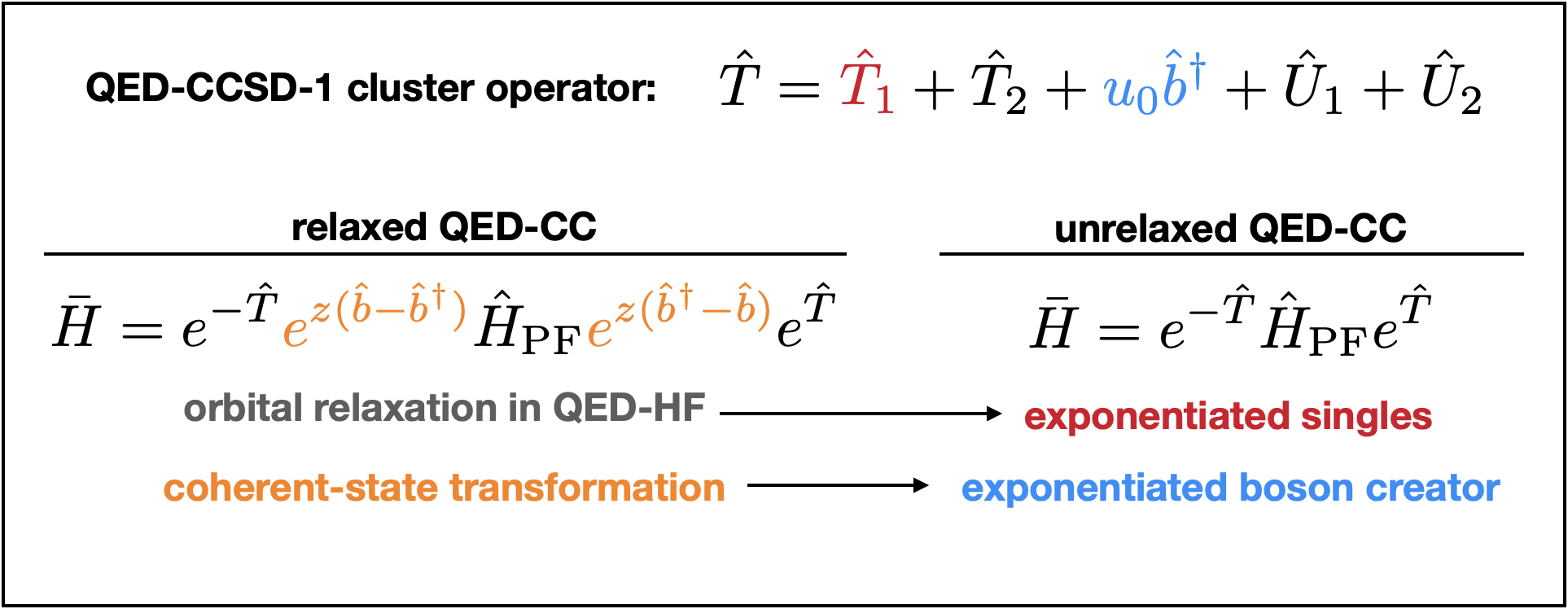}
\caption{The cluster operator and similarity-transformed Hamiltonian in relaxed and unrelaxed QED-CCSD-1 (QED-CC with up to single and double electronic excitations plus single photon creation operators).\cite{Koch20_041043} The single electron excitation and boson creation contributions to the cluster operator can account for the effects of orbital relaxation in QED-HF and the coherent-state transformation, respectively.}
\label{FIG:RELAXED_VS_UNRELAXED}
\end{figure}

The QED-CC wave function is defined as 
\begin{equation}
    \label{EQN:QED-CC-WFN}
    |\Psi_{\rm CC}\rangle = e^{\hat{T}}|\Phi_{0}\rangle
\end{equation}
where $\hat{T}$ is the cluster operator.  At the QED-CCSD-1 level of theory,\cite{Koch20_041043} $\hat{T}$ includes up to products of double electronic transitions and a single photon creation operator:
\begin{eqnarray}
    \label{EQN:T}
    \hat{T} &=& \sum_{ia} t_i^a \hat{a}^\dagger_a \hat{a}_i  + \frac{1}{4} \sum_{ijab} t_{ij}^{ab} \hat{a}^\dagger_a \hat{a}^\dagger_b \hat{a}_j \hat{a}_i \nonumber\\
    &+& u_0 \hat{b}^\dagger + \sum_{ia} u_i^a \hat{a}^\dagger_a \hat{a}_i \hat{b}^\dagger + \frac{1}{4} \sum_{ijab} u_{ij}^{ab} \hat{a}^\dagger_a \hat{a}^\dagger_b \hat{a}_j \hat{a}_i \hat{b}^\dagger
\end{eqnarray}
Here, $\hat{a}^\dagger$ and $\hat{a}$ represent fermionic creation and annihilation operators, respectively; the labels {\em i} / {\em j} and {\em a} / {\em b} refer to spin-orbitals that are occupied or unoccupied in the QED-HF reference wave function, respectively; and $t_i^a$, $t_{ij}^{ab}$, $u_0$, $u_i^a$, and $u_{ij}^{ab}$ are the cluster amplitudes. As mentioned above, in the case of unrelaxed QED-CC, the exponentiated single excitation operator, $e^{\hat{T}_1}$, can mimic the effects of cavity-induced orbital relaxation effects in relaxed QED-HF, and the term $e^{u_0\hat{b}^\dagger}$ is important for capturing the effects of the coherent-state transformation operator $\hat{U}_{\rm CS}$ itself (see Fig.~\ref{FIG:RELAXED_VS_UNRELAXED}).

The cluster amplitudes are determined using projective techniques, {\em i.e.}, by solving
\begin{eqnarray}
    \label{EQN:CC_ENERGY}
    \langle \mu^{\rm e} | \otimes \langle n  | e^{-\hat{T}} \hat{H}_{A} e^{\hat{T}}| 0 \rangle \otimes |0^{\rm e}\rangle  &=& \delta_{\mu 0} \delta_{n 0} E_{\rm CC}
\end{eqnarray}
Here, $\langle \mu^{\rm e}|$ and $\langle n|$ represent a determinant of spin-orbitals and a photon-number state with $n$ photons, respectively; $E_{\rm CC}$ is the energy associated with $|\Psi_{\rm CC}\rangle$; and the subscript $A$ refers to the type of Hamiltonian. For relaxed QED-CC, $\hat{H}_A = \hat{H}_{\rm CS}$; for unrelaxed QED-CC, $\hat{H}_A = \hat{H}_{\rm PF}$. At the QED-CCSD-1 level of theory, $\langle \mu^{\rm e}|$ can represent the reference or any singly- or doubly-substituted determinant of spin-orbitals, and $n$ can be zero or one.

\subsubsection{Excited-state QED-EOM-CC theory}

Given cluster amplitudes obtained by solving Eq.~\ref{EQN:CC_ENERGY},
excited states can be parametrized using the QED-EOM-CC formalism.\cite{Koch20_041043} {\color{black}The} left- and right-hand QED-EOM-CC wave functions {\color{black} are} defined by
\begin{eqnarray}
    \langle \tilde{\Psi}_I | = \langle \Phi_{0} | e^{-\hat{T}} \hat{L}_I\\
    | \Psi_I \rangle = \hat{R}_I e^{\hat{T}} | \Phi_{0}  \rangle
\end{eqnarray}
where the label {\em I} denotes the state.
At the QED-EOM-CCSD-1 level,\cite{Koch20_041043} the operators $\hat{L}_I$ and $\hat{R}_I$ take the form
\begin{eqnarray}
    \hat{L}_I &=&{^Il}_0 + \sum_{ai} {^Il}^i_a \hat{a}^\dagger_i\hat{a}_a + \frac{1}{2} \sum_{abij} {^Il}_{ab}^{ij} \hat{a}^\dagger_i \hat{a}^\dagger_j \hat{a}_b \hat{a}_a \nonumber\\
    &+&{^Im}_0 \hat{b} + \sum_{ai} {^Im}^i_a \hat{a}^\dagger_i\hat{a}_a \hat{b} + \frac{1}{2} \sum_{abij}{^Im}_{ab}^{ij} \hat{a}^\dagger_i \hat{a}^\dagger_j \hat{a}_b \hat{a}_a \hat{b}
\end{eqnarray}
and
\begin{eqnarray}
    \hat{R}_I &=&{^Ir}_0 + \sum_{ai} {^Ir}^a_i \hat{a}^\dagger_a\hat{a}_i + \frac{1}{2} \sum_{abij} {^Ir}_{ij}^{ab} \hat{a}^\dagger_a \hat{a}^\dagger_b \hat{a}_j \hat{a}_i \nonumber\\
    &+&{^Is}_0 \hat{b}^\dagger + \sum_{ai} {^Is}^a_i \hat{a}^\dagger_a\hat{a}_i \hat{b}^\dagger + \frac{1}{2} \sum_{abij} {^Is}_{ij}^{ab} \hat{a}^\dagger_a \hat{a}^\dagger_b \hat{a}_j \hat{a}_i \hat{b}^\dagger
\end{eqnarray} 
The $l$/$m$ and $r$/$s$  amplitudes are determined by solving left- and right-hand eigenvalue equations
\begin{equation}
    \langle \Phi_{0} |  \hat{L}_I \bar{H}  =  \langle \Phi_{0} | \hat{L}_I E_I
\end{equation}
and
\begin{equation}
    \bar{H} \hat{R}_I |\Phi_{0} \rangle = E_I \hat{R}_I |\Phi_{0} \rangle
\end{equation} 
where $\bar{H} = e^{-\hat{T}}\hat{H}_{A}e^{\hat{T}}$ is the similarity-transformed Hamiltonian, and $E_I$ represents the energy of the $I^{th}$ state. As in QED-CC, the choices $\hat{H}_A = \hat{H}_{\rm CS}$ and $\hat{H}_A = \hat{H}_{\rm PF}$ lead to relaxed and unrelaxed forms of QED-EOM-CC, respectively.

\subsection{Cavity QED Density Functional Theory}

A large body of literature describes quantum electrodynamical generalizations of DFT and TDDFT that differ in several aspects. First, for electronic degrees of freedom, some of these approaches represent the electronic density in real space,\cite{Rubio15_15285,Rubio18_992,Rubio19_2757,Narang20_094116,Varga22_194106} whereas others use atom-centered Gaussian basis functions.\cite{Shao21_064107,Shao22_124104,DePrince22_9303,Rubio22_094101} Second, photon degrees of freedom can be represented directly in real-space\cite{Tokatly13_233001, Rubio14_012508, Rubio2020_124119} or in Fock space\cite{Shao21_064107,Shao22_124104,Varga22_194106,DePrince22_9303,Rubio22_094101} (in a basis of photon-number states). Third, as with standard TDDFT, both real-time\cite{Tokatly13_233001,Varga22_194106} and linear-response\cite{Rubio19_2757,Narang20_094116,Shao21_064107,Shao22_124104,DePrince22_9303} formulations have been put forward. In this work, we consider linear-response QED-TDDFT formulated in terms of Gaussian basis functions and a Fock-space representation of the photon degrees of freedom. In analogy to the relaxed and unrelaxed QED-EOM-CC methods described above, we consider both relaxed and unrelaxed linear-response QED-TDDFT approaches that are equivalent to those described in Refs.~\citenum{Shao21_064107} and \citenum{DePrince22_9303}.

\subsubsection{Cavity QED Kohn-Sham DFT}

The QED-HF procedure outlined above can easily be adapted to the case of QED Kohn-Sham DFT. First, {\color{black}one can} map the QED-DFT ground-state onto a non-interacting state of the form of Eq.~\ref{EQN:QED_HF_WFN}, where $|0^{\rm e}\rangle$ now refers to a determinant of Kohn-Sham orbitals. Second, similar to the case of QED-HF, a ``relaxed'' Kohn-Sham determinant can be determined from an SCF procedure that makes use of the coherent-state Hamiltonian in Eq.~\ref{EQN:PFH_COHERENT_SCF}, with the energy augmented by a standard exchange-correlation functional. On the other hand, an ``unrelaxed'' QED-DFT state can be obtained from an SCF procedure that neglects the dipole self-energy contribution to Eq.~\ref{EQN:PFH_COHERENT_SCF}. Note that our formulations of relaxed and unrelaxed QED-DFT both ignore electron-photon correlation effects, such as those captured by the functionals described in Refs.~\citenum{Rubio15_093001, Rubio18_992, Rubio21_e2110464118,Flick22_143201}.

\subsubsection{Cavity QED Time-Dependent Density Functional Theory}

Excited states in QED-TDDFT are parametrized as
\begin{equation}
|\Psi_I\rangle = \hat{O}_I^\dagger | \Phi_0 \rangle
\end{equation}
with 
\begin{equation}
\hat{O}^\dagger_I = \sum_{ia} (X^I_{ia} \hat{a}^\dagger_a \hat{a}_i - Y^i_{ia}\hat{a}^\dagger_i\hat{a}_a) + M^I \hat{b}^\dagger -N^I \hat{b}
\end{equation}
In analogy to Rowe's equation of motion method,\cite{Rowe68_153} this parameterization leads to a generalized eigenvalue problem of the form 
\begin{widetext}
    \begin{equation}
        \label{EQN:QED_TDDFT_EQUATIONS}
        \begin{pmatrix}
            \bm{A+\Delta} &\bm{B+\Delta^\prime} & \bm{ g^\dagger} & \bm{ g^\dagger} \\
            \bm{B+\Delta^\prime} &\bm{A+\Delta} & \bm{ g^\dagger} & \bm{ g^\dagger}\\
            \bm{ g} & \bm{ g} &\bm{ \omega_{\rm cav}} &0  \\
            \bm{ g} & \bm{ g} &0 &\bm{ \omega_{\rm cav}}\\
        \end{pmatrix}
        \begin{pmatrix}
            X  \\
            Y \\
            M \\
            N\\
        \end{pmatrix}
        =\Omega
        \begin{pmatrix}
            \bm{1} &0 & 0 & 0 \\
            0 &\bm{-1} & 0 & 0\\
            0 & 0 &\bm{1} &0  \\
            0 & 0 &0 &\bm{-1}  \\
        \end{pmatrix}
        \begin{pmatrix}
            X  \\
            Y \\
            M \\
            N\\
        \end{pmatrix}
    \end{equation}
\end{widetext}
Here, ${\bm A}$ and ${\bm B}$ are the same matrices that arise in the usual random phase approximation (RPA) problem, {\em e.g.}, 
\begin{equation}
\langle \Phi_0 | [\hat{a}_i^\dagger \hat{a}_a, [\hat{H}_A, \hat{a}^\dagger_b \hat{a}_j]] | \Phi_0 \rangle  = ({\bm A} + {\bm \Delta})_{ai, bj}
\end{equation}
etc., with exchange contributions of ${\bm A}$ and ${\bm B}$ replaced/augmented by appropriate derivatives of the exchange-correlation functional for TDDFT. The symbols ${\bm \Delta}$ and ${\bm \Delta^\prime}$ represent dipole self-energy contributions of the form
\begin{eqnarray}
\label{EQN:DELTA}
\Delta_{ai,bj} = d_{ai} d_{jb} - d_{ab} d_{ij} \\
\label{EQN:DELTA_PRIME}
\Delta^\prime_{ai,bj} = d_{ai} d_{bj} - d_{aj} d_{ib}
\end{eqnarray}
where $d_{ai}$ is a dressed dipole integral
\begin{equation}
    d_{ai} = - \sum_{\xi \in \{x,y,z\}} \lambda_\xi \int \phi^*_a r_\xi \phi_{i} d\tau
\end{equation}
Here, $\phi$ is a Kohn-Sham orbital, $\lambda_\xi$ is a cartesian component of ${\bm{\lambda}}$, and $r_\xi$ is a cartesian component of the position vector [{\em e.g.}, for ${\mathbf{r}} = (x, y, z)$, $r_x$ = $x$]. For relaxed QED-TDDFT, $\hat{H}_A = \hat{H}_{\rm CS}$, and for unrelaxed QED-TDDFT, $\hat{H}_A = \hat{H}_{\rm PF}$. In order to recover the same equations as those used in the unrelaxed QED-TDDFT formalism of Ref.~\citenum{Shao21_064107}, one must also neglect the exchange contributions to ${\bm \Delta}$ and ${\bm \Delta}^\prime$ in Eqs.~\ref{EQN:DELTA} and \ref{EQN:DELTA_PRIME}.

\section{Computational Details}

The QED-TDDFT {\color{black} and QED-EOM-CCSD-1 methods were} implemented in \texttt{hilbert},\cite{hilbert} which is a plugin to the \textsc{Psi4}\cite{Sherrill20_184108} electronic structure package. Equations for the QED-CCSD-1 and QED-EOM-CCSD-1 were generated using a locally-modified version of \texttt{p$^\dagger$q},\cite{DePrince21_e1954709} which is a library for manipulating strings of second-quantized operators such as those that arise in coupled-cluster theory. All QED calculations used the 6-311++G** basis set with Cholesky-decomposed two-electron integrals and a tight decomposition threshold of 10$^{-12}$ E$_{\rm h}$. 
As mentioned in {\color{black}the Theory section}, QED-TDDFT calculations used standard density functional approximations from electronic structure theory that neglect electron-photon correlation effects. Geometries for all molecules were optimized at the DFT level of theory, using the 6-311++G** basis set, exact two-electron integrals, and the PBE0\cite{Barone99_6158} density functional. {\color{black}Excited-state calculations were carried out in the appropriate basis of $S_z$ = 0 determinants.}

The excited state potential energy curves (PECs) in all QED figures of this work were analyzed using \texttt{SuaveStateScanner},\cite{SuaveStateScanner} which assigns consistent labels to multiple states along PECs by enforcing the continuity of the excited-state energies and properties ({\em e.g.}, transition dipole moments, oscillator strengths, and norms of the excitation operators). Having consistent state labels greatly simplifies comparisons between the various QED approaches we use, particularly since all of the calculations in this work are performed without enforcing spatial symmetry.

\section{Results and Discussion}

In this Section, we analyze the ground- and excited-state energies of a series of diatomic molecules (molecular hydrogen, hydrogen fluoride, and lithium fluoride), coupled to a single-mode optical cavity.
We use bond lengths of 0.746 \AA, 0.918 \AA, and 1.582 \AA~for H$_2$, HF, and LiF, respectively, and the symmetry labels used to describe excited states correspond to the molecular axis oriented in the $z$-direction.

\subsection{Ground-state energies of relaxed and unrelaxed QED-CCSD-1}

We begin by considering the sensitivity of ground-state energies from QED-CCSD-1 to the treatment of cavity effects at the mean-field level.
That is, we wish to assess how well exponentiated singles and boson creation operators in unrelaxed QED-CCSD-1 can mimic the effects of orbital relaxation and the coherent-state transformation in relaxed QED-HF and QED-CCSD-1.
Table ~\ref{TABLE:ground_orbital_relaxation} provides ground-state energies from relaxed QED-CCSD-1 for several molecules coupled to a single-mode cavity with a coupling strength of $\lambda = 0.05$ atomic units (a comparable table for $\lambda = 0.1$ a.u.\ can be found in the Supporting Information).
The cavity mode is chosen to be polarized along the molecular axis (resonant with the following states: $1{}^1$B$_{1{\rm u}}$ for H$_2$, $2{}^1$A$_1$ for HF, and $3{}^1$A$_1$ for LiF) or perpendicular to the molecular axis (resonant with the following states:  $1{}^1$B$_{2{\rm u}}$ for H$_2$ or $1{}^1$B$_1$ for HF and LiF).
Also provided in Table~\ref{TABLE:ground_orbital_relaxation} are errors in unrelaxed QED-CCSD-1 energies with respect to the relaxed ones.
For H$_2$, we see that unrelaxed and relaxed QED-CCSD-1 agree to at least 10$^{-9}$ E$_{\rm h}$, but errors on the order of $10^{-4}$ E$_{\rm h}$ are observed for HF and LiF; the largest discrepancy between unrelaxed and relaxed QED-CCSD-1 energies is observed for LiF, with the cavity mode polarized along the molecular axis and resonant with the 1${}^1$A$_1$ $\to$ 3${}^1$A$_1$ transition ($\approx 0.06 \times 10^{-3}$ E$_{\rm h}$).
Considering the substantial coupling strength used ($\lambda = 0.05$ atomic units), the magnitudes of these differences suggest that exponentiated singles and boson creation operators do a reasonable job of capturing the effects of both orbital relaxation and the coherent-state transformation in relaxed QED-CCSD-1.

We have also evaluated errors in relaxed and unrelaxed QED-CCSD-1 energies when ignoring the exponentiated boson creation operator term, e$^{u_0\hat{b}^\dagger}$, in the QED-CCSD-1 cluster operator (labeled ``error w/o $u_0$'' in Table~\ref{TABLE:ground_orbital_relaxation}). For relaxed QED-CCSD-1, we see negligible energy deviations from full relaxed QED-CCSD-1; the largest deviations are on the order of 10$^{-6}$ E$_{\rm h}$.
On the other hand, this term is quite important for unrelaxed QED-CCSD-1, where energy errors as large as 0.008 E$_{\rm h}$ are observed.
The relative importance of the exponentiated boson creation operator in relaxed and unrelaxed QED-CCSD-1 is reflected in the value of $u_0$, which is also tabulated in Table~\ref{TABLE:ground_orbital_relaxation}.
We find that, when $u_0$ is non-zero, it can be more than an order of magnitude larger in the unrelaxed case.
We also note that, with the exception of one case (H$_2$ with the cavity mode resonant with the 1${}^1$B$_{1{\rm u}}$ state), $u_0$ is only non-zero when the cavity mode is polarized along the molecular axis.

\begin{figure}
    \centering
    \includegraphics{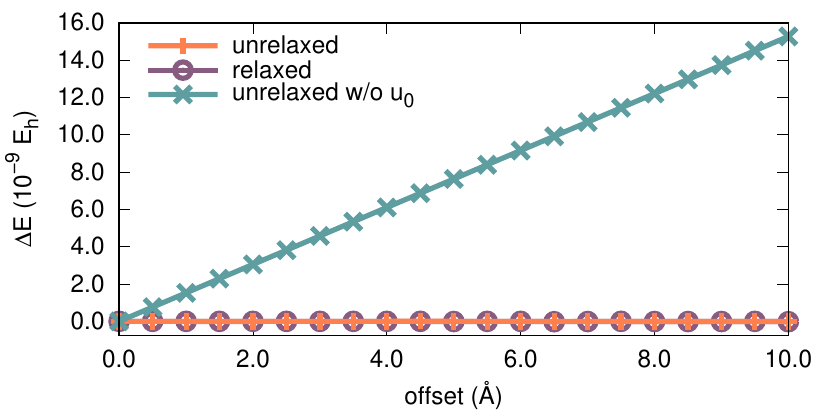}
    \caption{Origin dependence of the ground-state QED-CCSD-1 energy ($10^{-9}$E$_{\rm h}$) for hydrogen fluoride coupled to a cavity mode with a coupling strength of $\lambda = 0.05$. }
    \label{FIG:ORIGIN_INVARIANCE}
\end{figure}

Part of the motivation for the use of the coherent-state transformed Hamiltonian in relaxed QED-CCSD-1 is that it lends the origin invariance of QED-HF to the correlated problem.
On the other hand, an unrelaxed QED-CCSD-1 protocol should not be strictly origin invariant, although we expect the exponentiated boson creation operator to mitigate these effects.
Figure~\ref{FIG:ORIGIN_INVARIANCE} depicts how the energy from ground-state QED-CCSD-1 changes as calculations are carried out at various distances from the origin.
We consider hydrogen fluoride with a fixed bond length of 0.918 \AA~coupled to a cavity mode polarized along the molecular axis, resonant with the 2${}^1$A$_1$ state, and with a coupling strength of $\lambda = 0.05$ a.u. The distance along the $z$-axis in Fig.~\ref{FIG:ORIGIN_INVARIANCE} corresponds to the distance from the center of the bond to the origin, and the translation from the origin is carried out in the direction of the polarization of the cavity mode.
The change in the energy, $\Delta E$, corresponds to the difference between energies evaluated at the origin and away from it.
These data show that the relaxed QED-CCSD-1 energy is origin invariant, as expected.
Two sets of unrelaxed QED-CCSD-1 data are provided: one in which we include the exponentiated boson creation operator (labeled ``unrelaxed'') and one where we have neglected this term (labeled ``unrelaxed w/o $u_0$). We find that the $e^{u_0\hat{b}^\dagger}$ term is necessary for preserving the origin invariance of unrelaxed QED-CCSD-1;
ignoring this term introduces a small origin dependence in the energy (on the order of $10^{-9}$--$10^{-8}$ E$_{\rm h}$).
We note that Eqs.~\ref{EQN:COHERENT_TRANSFORM} and \ref{EQN:Z} show that the coherent-state transformation operator depends on the expectation value of the total QED-HF dipole moment, which should be origin invariant for neutral species.
By analogy, if $e^{u_0\hat{b}^\dagger}$ mimics the behavior of this term for unrelaxed QED-CCSD-1, $u_0$ itself should also be origin invariant; we have confirmed numerically that this is the case.

\begin{figure}
    \centering
    \includegraphics{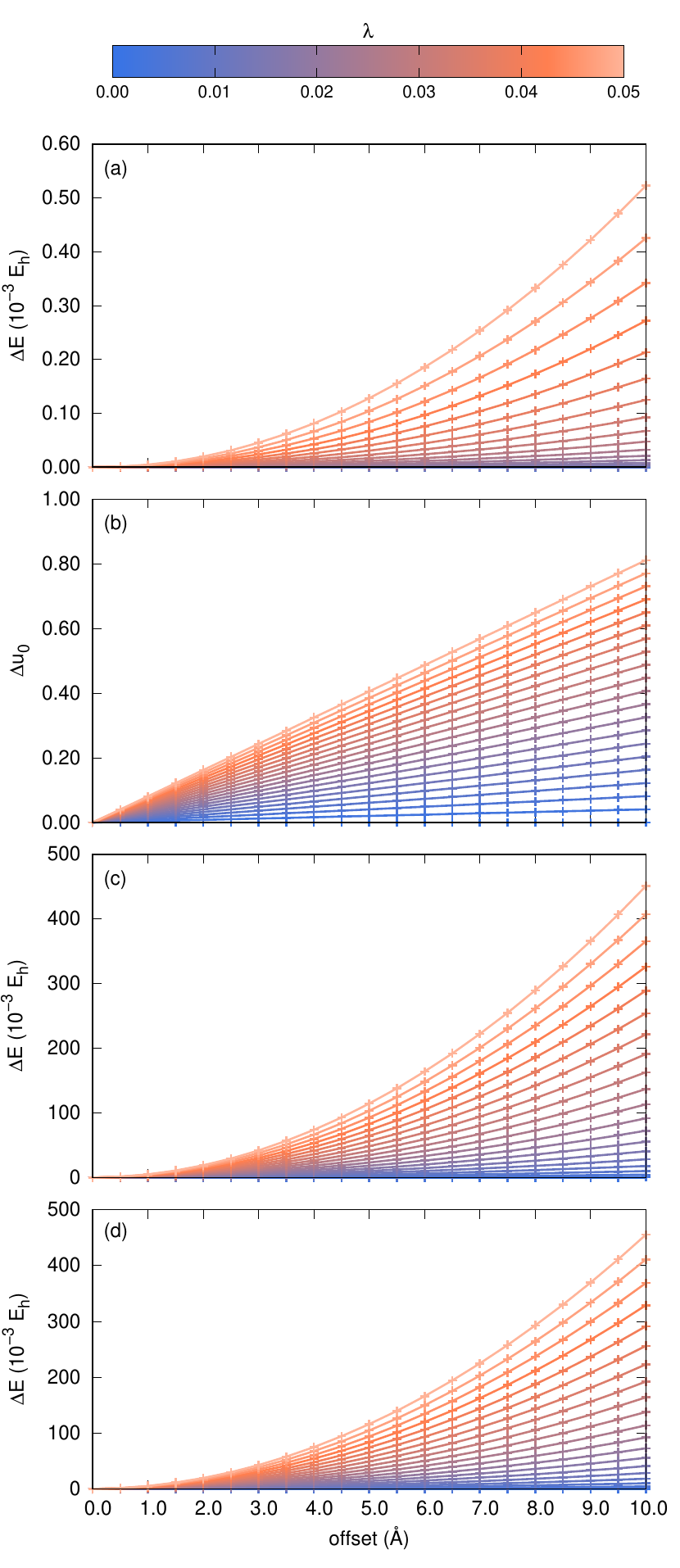}
    \caption{Origin dependence of unrelaxed QED-HF and QED-CCSD-1 at various coupling strengths ($\lambda$) and distances from the origin (10 \AA) for charged species (HF$^+$) with $\omega_{\rm cav}=0.675019$ E$_{\rm h}$. Panel (a) depicts the how the unrelaxed QED-CCSD-1 energy differs at a given offset from that at the origin. Panel (b) shows how $u_0$ in unrelaxed QED-CCSD-1 changes as the molecule is translated away from the origin. Panel (c) depicts similar information as panel (a), except that $u_0$ has been excluded from the cluster operator in unrelaxed QED-CCSD-1. 
    Panel (d) provides the difference between the unrelaxed QED-HF energy at a given offset and that at the origin.}
    \label{FIG:ORIGIN_INVARIANCE_CHARGED}
\end{figure}

Figure~\ref{FIG:ORIGIN_INVARIANCE_CHARGED} depicts a similar study for a charged species (HF$^+$, with an H--F distance of 0.918 \AA). In this case, the QED-HF dipole moment should depend on the placement of the molecule relative to the origin, and thus, we expect $u_0$ to also acquire an origin dependence in unrelaxed QED-CCSD-1.
The energy from relaxed QED-CCSD-1 is strictly origin invariant and is not shown.
For unrelaxed QED-CCSD-1, if we include the exponentiated boson operator, the energy does pick up a slight origin dependence; at 10 \AA~from the origin {\color{black} with $\lambda=0.05$ a.u.}, the energy differs from that at the origin by roughly 5 $\times$ 10$^{-4}$ E$_{\rm h}$ [see panel (a)].
On the other hand, without contributions from $u_0$, this energy difference grows to almost 0.5 E$_{\rm h}$ [see panel (c)].
As for $u_0$ itself, this quantity is origin invariant in the case of relaxed QED-CCSD-1 (not depicted), but it acquires a strong origin dependence in unrelaxed QED-CCSD-1 [see panel (b)].
The value of $u_0$ changes by roughly 0.8 when the molecule is translated by 10 \AA~from the origin {\color{black} at $\lambda=0.05$ a.u}.
As already mentioned, the origin dependence of $u_0$ is expected, as it mimics the coherent-state transformation; this transformation is defined by the mean-field dipole moment, which is strongly origin dependent for charged species.

Before moving on to discuss excited-states from unrelaxed and relaxed QED methods, we highlight the severe origin dependence of the energy for unrelaxed QED mean-field for cavity-coupled HF$^+$.  Panel (d) of Fig.~\ref{FIG:ORIGIN_INVARIANCE_CHARGED} depicts differences between the unrelaxed QED-HF energy evaluated at various distances from the origin relative to that computed at the origin. Clearly, the mean-field energy depends strongly on the choice of origin, and this dependence is of comparable magnitude to that which we observed for unrelaxed QED-CCSD-1 when ignoring $u_0$ [panel (c)]. This dependence is entirely due to the dipole self-energy contribution, and, since the dipole self energy term is treated in the same way in unrelaxed QED-DFT, that method also suffers from the same severe origin dependence.

\begin{table*}[!htb]
    \centering
    \caption{
        Ground-state energies ($E_{\rm h})$ from relaxed QED-CCSD-1 and absolute energy errors (10$^{-3}$ $E_{\rm h}$) from unrelaxed QED-CCSD-1,
        as well as from relaxed and unrelaxed QED-CCSD-1 calculations that ignore $u_0$. Also provided are $u_0$ values from relaxed and unrelaxed QED-CCSD-1 calculations.
        All calculations consider $\lambda=0.05$ atomic units.
    }
    \label{TABLE:ground_orbital_relaxation}
    \setlength{\tabcolsep}{0.2em}
    \begin{tabular*}{1.0\textwidth}{ cc@{\extracolsep{\fill}}ccccccc}
        \toprule
        & & & & \multicolumn{1}{c}{error} & \multicolumn{2}{c}{error w/o $u_0$} & \multicolumn{2}{c}{$u_0$}\\
        \cline{5-5} \cline{6-7} \cline{8-9}
        system & $\omega_{\rm cav}$ & resonance & relaxed & unrelaxed & relaxed & unrelaxed & relaxed & unrelaxed \\
        \midrule
        H$_2$ & $0.466751$ & 1$^1$B$_{1{\rm u}}$ &   $-1.167161$ & $0.000000$ & $0.000000$ & $0.000000$ &  $0.000000$ & $0.000000$ \\
        H$_2$ & $1.522218$ & 1$^1$B$_{2{\rm u}}$ &   $-1.167070$ & $0.000000$ & $0.000000$ & $0.000000$ &  $0.000000$ & $0.000000$ \\
        HF    & $0.531916$ & 2${}^1$A$_1$        & $-100.296930$ & $0.017471$ & $0.002001$ & $0.757147$ & $-0.001815$ & $0.037626$ \\
        HF    & $0.375022$ & 1${}^1$B$_1$        & $-100.296806$ & $0.015717$ & $0.000054$ & $0.015663$ &  $0.000000$ & $0.000000$ \\
        LiF   & $0.308401$ & 3${}^1$A$_1$        & $-107.233438$ & $0.062602$ & $0.005998$ & $8.095084$ &  $0.003957$ & $0.162599$ \\
        LiF   & $0.232119$ & 1${}^1$B$_1$        & $-107.220994$ & $0.039174$ & $0.000182$ & $0.038991$ &  $0.000000$ & $0.000000$ \\
        \bottomrule
    \end{tabular*}
\end{table*}

\subsection{Excitation energies of relaxed and unrelaxed QED-TDDFT and QED-EOM-CCSD-1}

We now consider the effects that orbital relaxation and the coherent-state transformation have on excitation energies derived from QED-TDDFT and QED-EOM-CCSD-1. 
We have performed relaxed and unrelaxed QED-TDDFT and QED-EOM-CCSD-1 calculations on cavity-coupled molecules with the cavity mode resonant with the following states:  {\color{black} 1$^1$B$_{1{\rm u}}$ and  1$^1$B$_{2{\rm u}}$ for H$_2$, 2${}^1$A$_1$ and 1${}^1$B$_1$ for HF, and 3${}^1$A$_1$ and 1${}^1$B$_{1}$ for LiF.}
We compare the computed excitation energies of the relaxed and unrelaxed formulations of QED-TDDFT to QED-EOM-CCSD-1.
Specifically, Figs.~\ref{FIG:H2_PEC}, \ref{FIG:HF_PEC}, and \ref{FIG:LIF_PEC} depict changes in excitation energies for H$_2$, HF, and LiF, respectively, as the cavity coupling strength is increased.
Each figure assumes the following format:
\begin{itemize}
    \item Panels on the left and right correspond to calculations for which the polarization of the cavity mode was parallel to the molecular axis or perpendicular to it, respectively, with the cavity frequency resonant with the appropriate cavity-free transition (see above for the specific states we target).
    \item The top panels depict the vertical excitation energies (VEE) for several states, shifted by the cavity frequency ($\omega_{\rm cav}$), as a function of the cavity coupling strength ($\lambda$).
    \item Relaxed and unrelaxed QED-TDDFT curves are purple and orange, respectively, while the green and black curves correspond to relaxed and unrelaxed QED-EOM-CC. 
    \item Solid lines correspond to the polariton states, while dashed lines correspond to non-resonant states that are nearby in energy.
    For clarity, QED-EOM-CC states with significant double electronic transition character are not shown.
    \item The middle panels show how Rabi splittings from relaxed QED-TDDFT, unrelaxed QED-TDDFT, and unrelaxed QED-EOM-CC deviate from those from relaxed QED-EOM-CC ($\Delta \Omega_R$).
    \item The bottom panels present the deviation from relaxed QED-EOM-CC Rabi splittings as a percentage.
\end{itemize}

\subsubsection{Molecular hydrogen}
\begin{figure*}
    \centering
    \includegraphics[width=\linewidth]{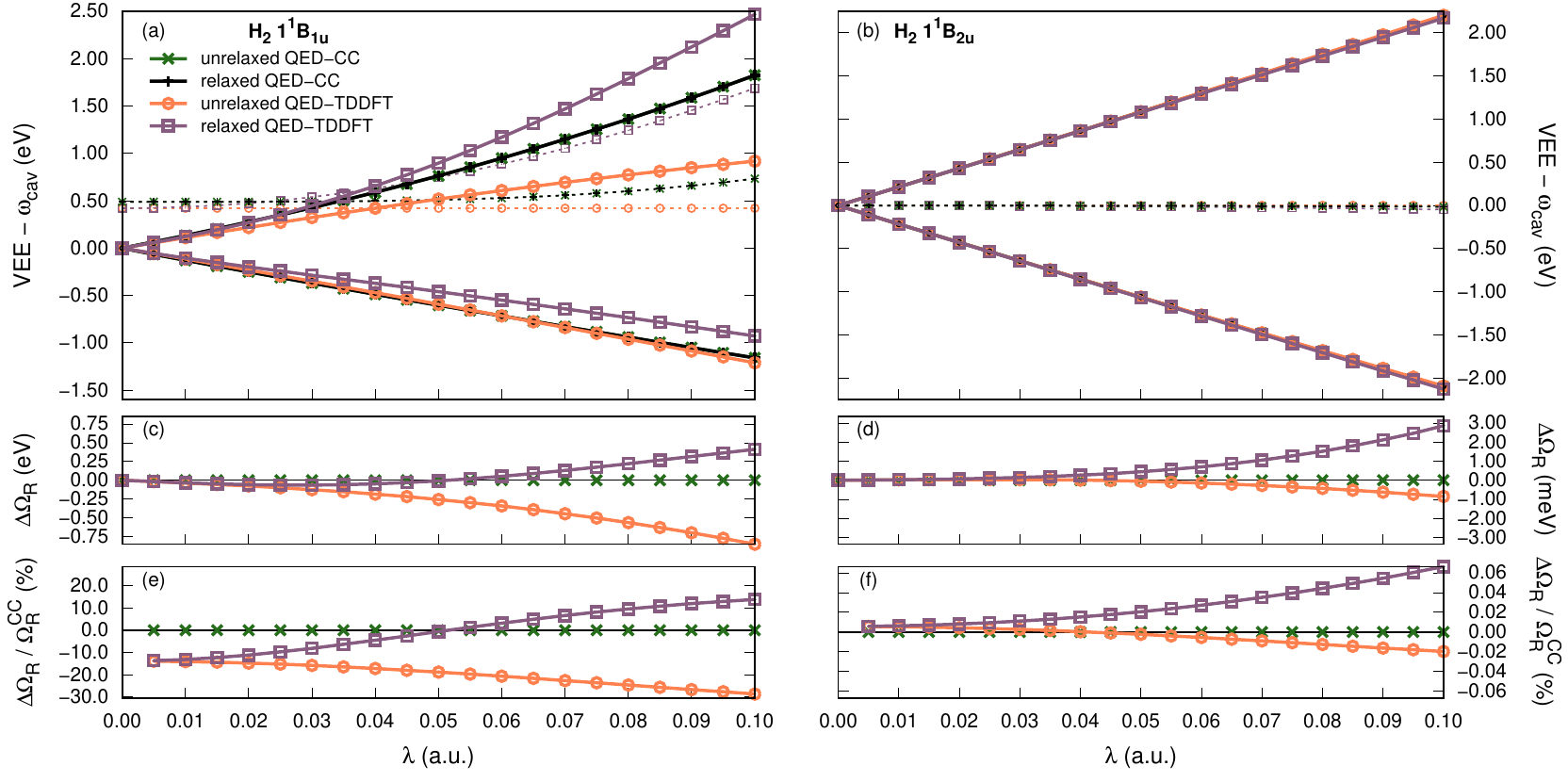}
    \caption{
        Excitation energies for H$_2$ when coupling a cavity mode to the (a) {\color{black}1$^1$B$_{1{\rm u}}$} and (b) {\color{black}1$^1$B$_{2{\rm u}}$} (b) states, using relaxed QED-TDDFT (purple), unrelaxed QED-TDDFT (orange), unrelaxed QED-EOM-CCSD-1 (green), and relaxed QED-EOM-CCSD-1 (black).
        The dashed lines correspond to non-resonant excited states [2${}^1$A$_{\rm g}$ in panel (a) and {\color{black}1$^1$B$_{3{\rm u}}$} in panel (b)], while the solid lines correspond to the polariton states.
        Panels (c) and (d) show the difference between the Rabi splittings from relaxed QED-EOM-CCSD-1 to relaxed/unrelaxed QED-TDDFT and unrelaxed QED-EOM-CCSD-1, and panels (e) and (f) depict these differences as a percentage.
    }
    \label{FIG:H2_PEC}
\end{figure*}

Figure~\ref{FIG:H2_PEC} illustrates how the excited-state landscape of cavity-coupled H$_2$ changes with the coupling strength, for cavity modes that are resonant with the (a) {\color{black}1$^1$B$_{1{\rm u}}$} and (b) {\color{black}1$^1$B$_{2{\rm u}}$} states of cavity-free H$_2$. First, excitation energies from relaxed and unrelaxed QED-EOM-CC are indistinguishable on the scale of this figure{\color{black}, but they are not numerically identical. In this case, QED-EOM-CCSD-1 is equivalent to the full CI in the electronic space, so we expect the approach to be invariant to cavity-induced orbital relaxation effects. It is also equivalent to the full CI in the photon space, {\em if the photon space is truncated at two photon number states} (0 and 1). This qualifying statement is important, for the following reason. The coherent-state transformation of the Hamiltonian should preserve the spectrum of the Hamiltonian, but the spectrum is only preserved in the limit that the photon space is complete. Indeed, we do not see exact numerical agreement between relaxed and unrelaxed QED-EOM-CCSD-1 for this reason (differences on the order of 10$^{-6}$ E$_{\rm h}$ are observed for $\lambda = 0.1$ a.u.; see Supporting Information). Nonetheless, relaxed and unrelaxed QED-EOM-CC results are nearly indistinguishable in this case.} 

Curves corresponding to polariton formation involving the {\color{black} 1$^1$B$_{1{\rm u}}$} state reveal significant differences between QED-EOM-CC and both relaxed and unrelaxed QED-TDDFT.
In the case of relaxed QED-TDDFT, both the Rabi splitting (the difference in energy between the upper and lower polariton states, $\Omega_R$) and the energies of the states that are not resonant with the cavity mode are more sensitive to cavity effects in the strong coupling limit than the same quantities derived from QED-EOM-CC calculations; this is a general trend we observe for all systems considered in this work.
For a cavity mode resonant with the {\color{black}1$^1$B$_{1{\rm u}}$} state, we can see that the $\lambda$-dependence of the lower polariton state energy derived from QED-EOM-CC is in better agreement with unrelaxed QED-TDDFT than with relaxed QED-TDDFT. On the other hand, these three methods predict noticeably different trends in the $\lambda$-dependence of the upper polariton state. The shift in the upper polariton energy from unrelaxed QED-TDDFT is too small, and that from relaxed QED-TDDFT is too large; QED-EOM-CC splits the difference. In the large-$\lambda$ limit, Figs.~\ref{FIG:H2_PEC}(c) and (e) indicate that relaxed QED-TDDFT provides better Rabi splittings than unrelaxed QED-TDDFT, given QED-EOM-CC results as a reference. The maximum deviation between relaxed QED-TDDFT and QED-EOM-CC Rabi splittings is $0.41$~eV ($13.8\%$) at $\lambda = 0.1$, while  unrelaxed QED-TDDFT and QED-EOM-CC splittings differ by $-0.85$~eV ($-28.6\%$) at the same coupling strength. The $\lambda$-dependence of the non-resonant states [indicated by dashed lines in Fig.~\ref{FIG:H2_PEC}(a)] is also interesting. Here, we see the excitation energy of the non-resonant state is mostly unaffected by the cavity mode in the case of unrelaxed QED-TDDFT. On the other hand, both relaxed QED-TDDFT and QED-EOM-CC predict an increase in the excitation energy, and QED-TDDFT exaggerates this effect. We conclude that, because this state does not directly interact to the cavity mode, these changes with increasing coupling strength must stem from cavity-induced changes to the ground state. 

The curves for polariton formation with the {\color{black}1$^1$B$_{2{\rm u}}$ state of H$_2$ are depicted in the right-hand panels of Fig.~\ref{FIG:H2_PEC}. 
As was seen in panels (a), (c), and (e), relaxed and unrelaxed QED-EOM-CC display nearly identical $\lambda$ dependence, which is not surprising, given that QED-EOM-CCSD-1 is equivalent to the full CI (within the truncated photon space).
Unlike in panels (c) and (e),} Rabi splittings from unrelaxed and relaxed QED-TDDFT both agree well with those from QED-EOM-CC [panels (d) and (f)], with maximum deviations on the order of only $10^{-3}$ eV.
Unrelaxed QED-TDDFT displays slightly better agreement with QED-EOM-CC than relaxed QED-TDDFT, but the deviations from QED-EOM-CC overall are small for both QED-TDDFT variants.

\subsubsection{Hydrogen fluoride}

\begin{figure*}
    \centering
    \includegraphics[width=\linewidth]{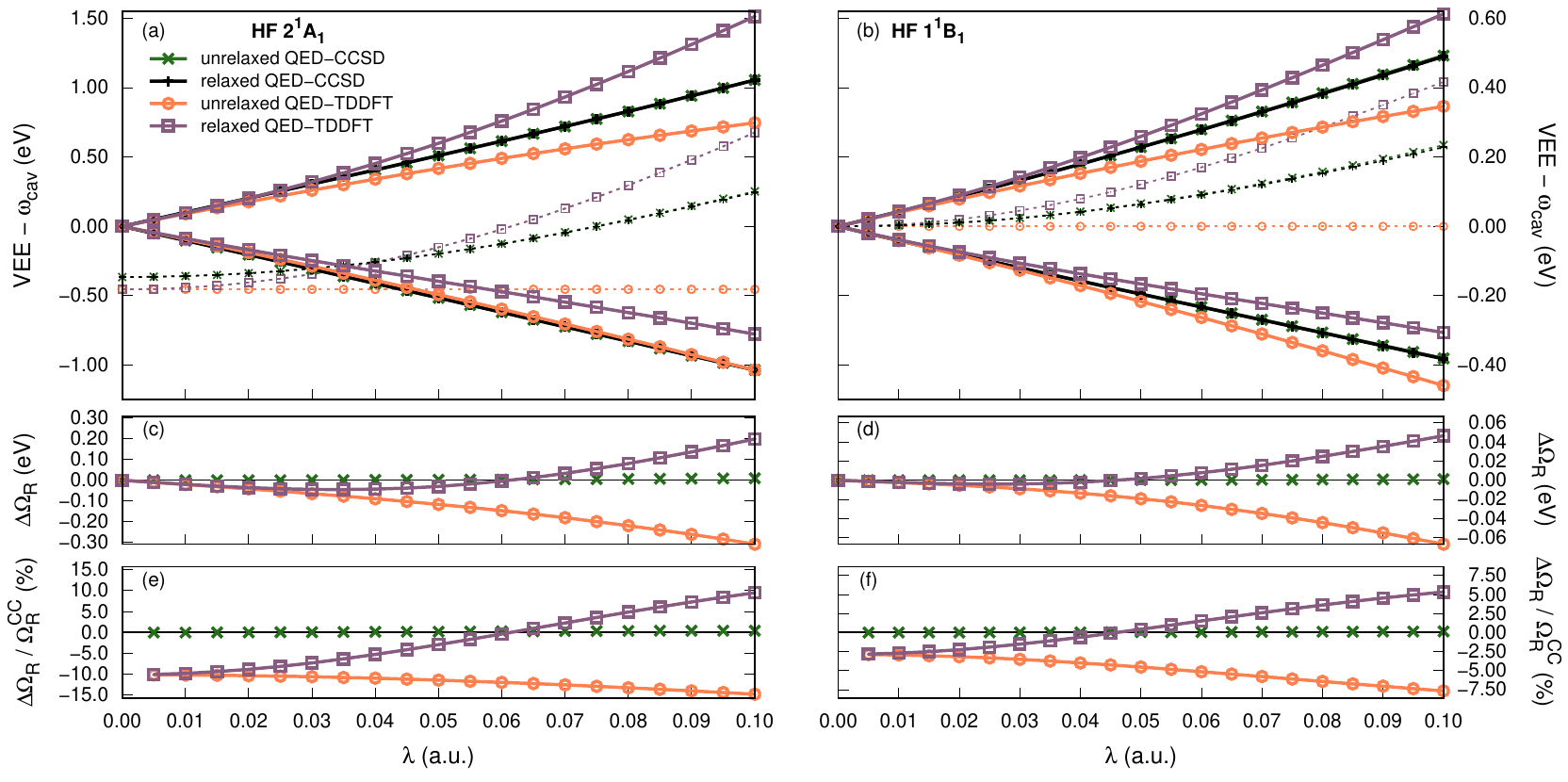}
    \caption{
        Excitation energies for HF when coupling a cavity mode to the (a) {\color{black}2$^1$A$_1$} and (b) {\color{black}1$^1$B$_{1}$} states, using relaxed QED-TDDFT (purple), unrelaxed QED-TDDFT (orange), unrelaxed QED-EOM-CCSD-1 (green), and relaxed QED-EOM-CCSD-1 (black).
        The dashed lines correspond to non-resonant excited states[{\color{black}2$^1$B$_1$} in panel (a) and {\color{black}1$^1$B$_{2}$} in panel (b)], while the solid lines correspond to the polariton states.
        Panels (c) and (d) show the difference between the Rabi splittings from relaxed QED-EOM-CCSD-1 to relaxed/unrelaxed QED-TDDFT and unrelaxed QED-EOM-CCSD-1, and panels (e) and (f) depict these differences as a percentage.
    }
    \label{FIG:HF_PEC}
\end{figure*}

Figure \ref{FIG:HF_PEC} provides similar data as Fig.~\ref{FIG:H2_PEC}, but for cavity-coupled hydrogen fluoride. The left panels correspond to calculations with the cavity mode polarized along the molecular axis and the cavity mode frequency resonant with the {\color{black}2$^1$A$_{1}$} state of cavity-free HF. The right panels consider a cavity mode polarized perpendicular to the molecular axis and resonant with the {\color{black}1$^1$B$_{1}$} state of isolated HF. The left panels show similar behavior as depicted in the left panels of Figure~\ref{FIG:H2_PEC} for H$_2$. First, relaxed and unrelaxed QED-EOM-CC results are indistinguishable. Second, the $\lambda$-dependence of the lower polariton from unrelaxed QED-TDDFT agrees well with that from QED-EOM-CC, while relaxed QED-TDDFT appears to underestimate this dependence. Third, unrelaxed and relaxed QED-TDDFT predict a $\lambda$-dependence for the upper polariton state that is too small or too large, as compared to that from QED-EOM-CC, respectively. In terms of the Rabi splitting [panels (c) and (e)], relaxed QED-TDDFT again provides a better description than unrelaxed QED-TDDFT, given relaxed QED-EOM-CC as the reference. Here, relaxed and unrelaxed QED-TDDFT Rabi splittings deviate from those of relaxed QED-EOM-CC by at most $0.20$~eV ($9.5\%$) and $-0.31$~eV ($-14.8\%$), respectively. The magnitudes of these deviations are smaller than in the case of H$_2$ in the left panels of Fig.~\ref{FIG:H2_PEC} above, which could simply reflect the smaller magnitude of the Rabi splitting itself for the {\color{black}2$^1$A$_{1}$} state of HF, relative to the splitting for the {\color{black}1$^1$B$_{1{\rm u}}$} state in H$_2$ (see Supporting Information). Indeed, the oscillator strength for the {\color{black}2$^1$A$_{1}$} state of HF ($0.1869$) is much smaller than that for the {\color{black}1$^1$B$_{1{\rm u}}$} state of H$_2$ ($0.3069$), which is consistent with the relative Rabi splittings. Lastly, as was observed in Fig.~\ref{FIG:H2_PEC}(a), the excitation energies for the non-polariton states in Fig.~\ref{FIG:HF_PEC}(a) pick up a $\lambda$-dependence in the case of both relaxed and unrelaxed QED-EOM-CC and for relaxed QED-TDDFT, with a more pronounced dependence for QED-TDDFT. On the other hand, unrelaxed QED-TDDFT predicts that these excitation energies are unaffected by the presence of the cavity.

The curves in the right panel of Fig.~\ref{FIG:HF_PEC} depict the $\lambda$-dependence of excitation energies and Rabi splittings when the cavity mode is resonant with the {\color{black}1$^1$B$_{1}$} state of isolated HF. The same qualitative observations for the left panels Fig.~\ref{FIG:HF_PEC} apply here, with the exception that the $\lambda$-dependence of the lower polariton is not well-reproduced by unrelaxed QED-TDDFT. Note also that the behavior here differs somewhat from the case of the cavity mode polarized perpendicular to the {\color{black}1${}^1$B$_{2{\rm u}}$} of isolated H$_2$. In that case, all QED approaches provided comparable results, whereas, here, relaxed QED-TDDFT does a better job of reproducing the $\lambda$-dependence of the Rabi splittings predicted by QED-EOM-CC [panels (d) and (f)]; 
relaxed and unrelaxed QED-TDDFT Rabi splittings deviate from relaxed QED-EOM-CC splittings by at most $0.046$~eV ($5.3\%$) and $-0.067$~eV ($-7.7\%$), respectively. The data in panel (b) also demonstrate that relaxed QED-TDDFT captures the same qualitative $\lambda$-dependence of the non-resonant state predicted by relaxed and unrelaxed QED-EOM-CC (albeit somewhat exaggerated by QED-TDDFT), whereas unrelaxed QED-TDDFT does not.

\subsubsection{Lithium fluoride}

\begin{figure*}
    \centering
    \includegraphics[width=\linewidth]{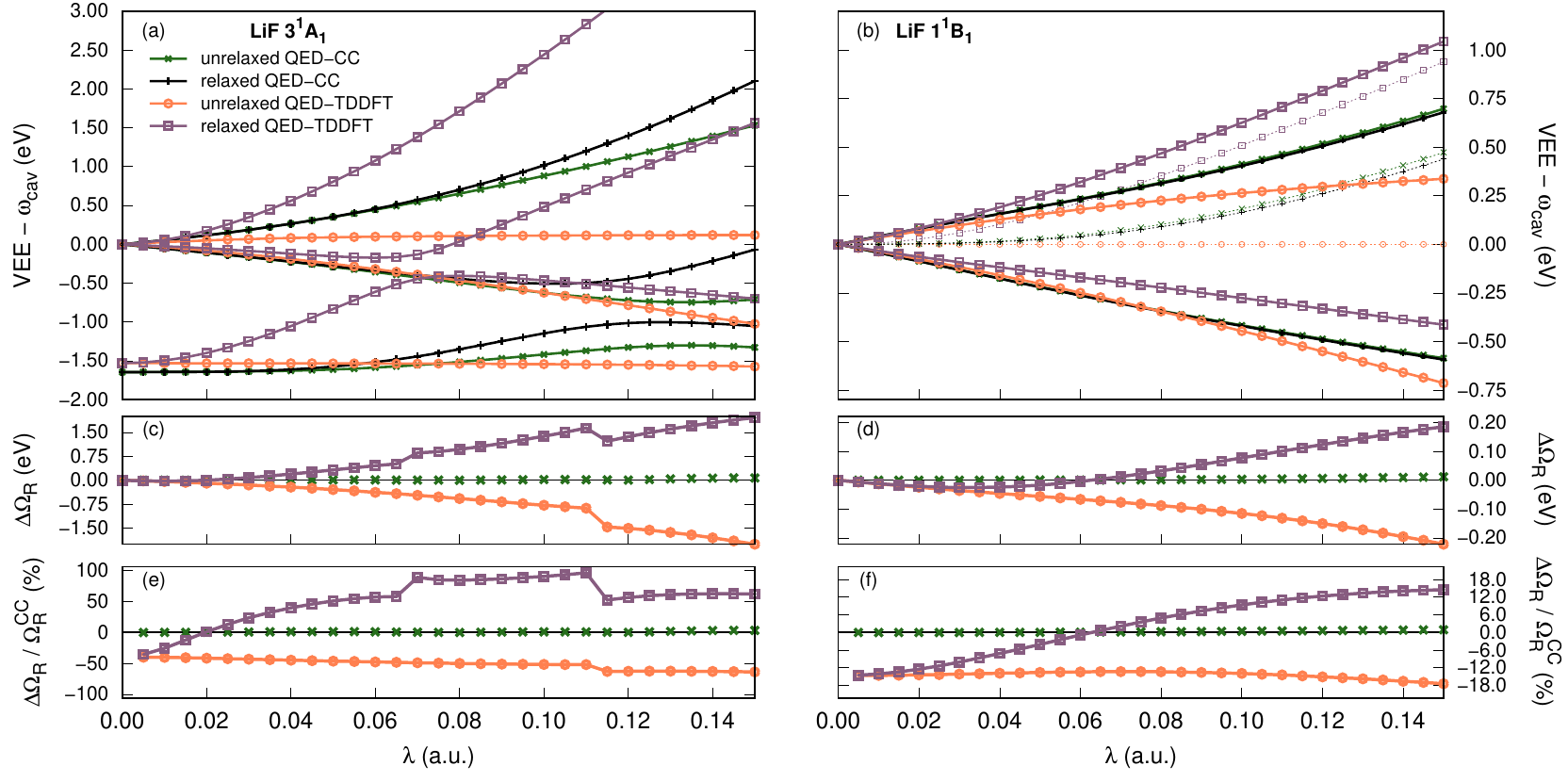}
    \caption{
        Excitation energies for LiF when coupling a cavity mode to the (a) {\color{black}3$^1$A$_1$} and (b) {\color{black}1$^1$B$_{1}$} states, using relaxed QED-TDDFT (purple), unrelaxed QED-TDDFT (orange), unrelaxed QED-EOM-CCSD-1 (green), and relaxed QED-EOM-CCSD-1 (black). 
        In panel (a), the curves at $\lambda = 0.00$ a.u.~correspond to the 2$^1$A$_1$ and {\color{black}3$^1$A$_1$} states of LiF.
        The dashed lines in panel (b) correspond to a non-resonant excited state [{\color{black}1$^1$B$_{2}$}], while the solid lines are the polariton states formed from coupling the cavity mode to the {\color{black}1$^1$B$_{1}$} state.
        Panels (c) and (d) show the difference between the Rabi splittings from relaxed QED-EOM-CCSD-1 to relaxed/unrelaxed QED-TDDFT and unrelaxed QED-EOM-CCSD-1, and panels (e) and (f) depict these differences as a percentage.
    }
    \label{FIG:LIF_PEC}
\end{figure*}

Finally, we come to the case of lithium fluoride. Figure \ref{FIG:LIF_PEC} illustrates the $\lambda$-dependence of the excitation energies and Rabi splittings from QED-TDDFT and QED-EOM-CC, for a cavity mode polarized along the molecular axis and resonant with the {\color{black}3$^1$A$_1$} of cavity-free LiF (left panels) and for a cavity mode polarized perpendicular to the molecular axis and resonant with the {\color{black}1$^1$B$_1$} of the isolated molecule (right panels). We note that {\color{black}3$^1$A$_1$} is the second bright ${}^1$A$_1$ state of LiF, whereas, in the previous examples, we had tuned the cavity to the lowest-energy bright state of the given symmetry. We note that we have not depicted non-resonant excited states in Fig.~\ref{FIG:LIF_PEC}(a) aside from the 2$^1$A$_1$ state for the sake of clarity {\color{black}(these other states have incompatible spatial or spin symmetry to couple directly to the cavity mode)}. That said, for such states, we observe the same trends as have been discussed in the context of the other systems; QED-EOM-CC and relaxed QED-TDDFT predict a $\lambda$-dependence in these states that is exaggerated by QED-TDDFT, and unrelaxed QED-TDDFT fails to capture this effect.

In Fig.~\ref{FIG:LIF_PEC}(a), we find that we can induce some interesting behavior by tuning to the second {\color{black} A$_1$ symmetry} bright state of LiF {\color{black}(3$^1$A$_1$)}, which leads to strong interactions between the lower polariton state and the first bright state (2$^1$A$_1$) at larger coupling strengths. Specifically, Fig.~\ref{FIG:LIF_PEC}(a) demonstrates a coupling-strength-induced avoided crossing between these states that appears at coupling strengths of roughly $\lambda = 0.070$, $\lambda = 0.115$, and $\lambda = 0.135$ a.u.~when modeling the system with relaxed QED-TDDFT, relaxed QED-EOM-CC, and unrelaxed QED-EOM-CC, respectively.  The avoided crossing appears at smaller $\lambda$ values for relaxed QED-TDDFT than for relaxed QED-EOM-CC because, as in the previous cases, relaxed QED-TDDFT exaggerates the $\lambda$-dependence of the states that are not resonant with the cavity mode. Notably, to a coupling strength of $\lambda = 0.15$ a.u., we do not observe this avoided crossing feature in unrelaxed QED-TDDFT because it fails to capture the $\lambda$-dependence of the 2$^1$A$_1$ state. Also noteworthy is that this example is the first instance where we observe appreciable differences between relaxed and unrelaxed QED-EOM-CC. Results from these methods are similar for coupling strengths up to $\lambda = 0.05$ a.u.~but differ for larger coupling strengths. These differences may not be terribly important in practical applications, though, given that these large coupling strengths correspond to effective cavity mode volumes that are significantly smaller than 1 nm$^3$. For example, $\lambda = 0.1$ a.u.~corresponds to an effective mode volume of less than 0.2 nm$^3$, which is smaller than any reported experimentally obtained value of which we are aware.

One interesting aspect of the avoided crossing is that the character of the lower polariton state is transferred to the lower-energy state beyond the avoided crossing, which points to potential ambiguities in designating one and only one state as an upper or lower polariton state in systems with dense energy manifolds. 
In order to compare Rabi splittings from each method before and after the avoided crossing, we simply take the point at which the energy gap between the states is the smallest as the point at which the crossing occurs;
the kinks observed in the curves depicted in panels (c) and (e) of Fig.~\ref{FIG:LIF_PEC} correspond to these points. 
These details aside, we find that neither unrelaxed QED-TDDFT nor relaxed QED-TDDFT do a particularly good job of reproducing the $\lambda$-dependence or Rabi splittings from relaxed QED-EOM-CC; in particular, QED-TDDFT Rabi splittings differ from relaxed QED-EOM-CC ones by roughly 2 eV at $\lambda$ = 0.15 a.u.
These deviations are much larger than those observed for other molecules, which reflects the complexity of the excited-state energy landscape of LiF and calls into question the reliability of either form of QED-TDDFT in this case. We also note that there are smaller $\lambda$ values for which unrelaxed QED-TDDFT Rabi splittings appear to be the better ones, relative to QED-EOM-CC. That said, the trends in panel (a) suggest that QED-TDDFT does a better job of reproducing qualitative properties of relaxed QED-EOM-CC when the procedure accounts for the effects of orbital relaxation and the coherent-state transformation. 
Aside from the poor behavior of QED-TDDFT, perhaps the most notable difference here as compared to the earlier examples is the discrepancy between relaxed and unrelaxed QED-EOM-CC. The Rabi splittings from these methods differs by as much as 72.2 meV (3.32\%) at $\lambda=0.15$, which is larger by an order of magnitude than other systems in this study (see Supporting Information). The $\lambda$-dependence for the 2$^1$A$_1$ state is also underestimated by unrelaxed EOM-QED-CC, which shifts the avoided crossing from  $\lambda=0.115$ a.u.~to $\lambda = 0.135$ a.u.


Lastly, we consider LiF coupled to a cavity mode polarized perpendicular to the molecular axis and resonant with the {\color{black}1$^1$B$_1$} state of the isolated molecule (right panels of Fig.~\ref{FIG:LIF_PEC}). In this case, the general trends are similar to what was observed when coupling a cavity mode to the {\color{black}1$^1$B$_1$} state of HF (right panels of Fig.~\ref{FIG:HF_PEC}). Again, ignoring orbital relaxation and the coherent-state transformation in QED-TDDFT decreases the ability of QED-TDDFT to reproduce relaxed QED-EOM-CC Rabi splittings; at $\lambda = 0.15$ a.u., the Rabi splitting from relaxed QED-TDDFT differs from the relaxed QED-EOM-CC value by $0.19$~eV ($14.6$\%); this deviation increases to $-0.22$~eV ($-17.4$\%) for unrelaxed QED-TDDFT. Also, as observed previously, unrelaxed QED-TDDFT fails to capture the $\lambda$-dependence of states other than the upper and lower polariton states; this dependence is captured by relaxed QED-TDDFT, but the sensitivity of these non-resonant states to the cavity is overestimated, relative to QED-EOM-CC. We also note that, as was observed for polariton formation with the {\color{black}3$^1$A$_1$} state, small differences between relaxed and unrelaxed QED-EOM-CC methods emerge at coupling strengths larger than $\lambda = 0.05$ a.u.

\section{Conclusions}

Recent intriguing experiments demonstrating the ability to manipulate chemical transformations via vacuum field fluctuations and polariton formation have inspired the development of several generalizations of standard electronic structure methods ({\em e.g.}, coupled cluster theory, density functional theory, configuration interaction, etc.) for the polariton problem. In this work, we have explored the numerical consequences of some formal aspects of QED-DFT, QED-TDDFT, QED-CCSD-1, and QED-EOM-CCSD-1. Specifically, we began by investigating the sensitivity of ground-state energies from QED-CCSD-1 to the treatment of cavity effects at the mean-field level. We have found that the inclusion of exponentiated single electron transitions and boson creation operators in QED-CCSD-1 makes the approach robust with respect to the inclusion or exclusion of cavity effects in the underlying QED-HF calculation; numerically, these terms do a good job of mimicking the effects of orbital relaxation and the coherent-state transformation, respectively. Exponentiated boson creation operators are particularly important for maintaining (or nearly maintaining, in the case of charged species) origin invariance in unrelaxed QED-CCSD-1. On the other hand, unrelaxed mean-field approaches display severe origin dependence for charged species, which arises from the dipole self-energy contribution to the energy.

Beyond the ground state, we have also assessed the role that cavity effects at the mean-field level can have on excited states computed using QED-TDDFT and QED-EOM-CCSD-1. Several key details bear repeating. First, for the most part, excitation energies computed from unrelaxed and relaxed QED-EOM-CC are similar; Rabi splittings differ by less than 9.3 meV (or less than 0.45\%) in all cases considered in this work, except near the avoided-crossing for the {\color{black}3${}^1$A$_1$} state of LiF which reaches an error of 72.2 meV (3.32\%) at the extreme case of $\lambda=0.15$ a.u.~(see Supporting Information).
Second, QED-EOM-CC and relaxed QED-TDDFT predict that the energies of electronic states that are not resonant with the cavity mode can be significantly perturbed in the strong coupling limit, and relaxed QED-TDDFT exaggerates this effect. On the other hand, unrelaxed QED-TDDFT fails to predict any $\lambda$ dependence in non-resonant states. Third, Rabi splittings from QED-EOM-CC are more closely reproduced by relaxed QED-TDDFT than by unrelaxed QED-TDDFT. In the large coupling limit, relaxed QED-TDDFT tends to overestimate the Rabi splitting, while unrelaxed QED-TDDFT underestimates this quantity. Lastly, the proximity of multiple bright states having the appropriate symmetry to interact with the cavity mode can lead to complex spectral features; specifically, we have located a coupling-strength-induced avoided crossing in LiF between the lower polariton (formed from the admixture of the cavity mode and the {\color{black}$3{}^1$A$_1$} state) and the $2{}^1$A$_1$ state. Of the methods studied, unrelaxed QED-TDDFT is the least capable of describing this phenomenon, because it fails to capture the $\lambda$-dependence of the non-resonant $2{}^1$A$_1$ state. Given these observations, we caution against the use of unrelaxed QED-TDDFT.
\vspace{0.5cm}

{\bf Supporting Information} Relaxed and unrelaxed QED-CCSD-1 energies for ground states of molecular hydrogen, hydrogen fluoride, and lithium fluoride; Rabi splittings from relaxed QED-EOM-CCSD-1 for these same molecules; deviations in Rabi splittings computed using relaxed and unrelaxed QED-EOM-CCSD-1.

\vspace{0.5cm}

\noindent {\bf Acknowledgments Information:}
This material is based upon work supported by the National Science Foundation under Grant No.~CHE-2100984.\\ 

\bibliography{Journal_Short_Name.bib,cqed.bib,cc.bib,deprince.bib,packages.bib, other_ref.bib}

\end{document}